\documentclass[aps,prb,twocolumn,superscriptaddress,notitlepage,nofootinbib,longbibliography, colorlinks=true,10pt]{revtex4-2}
\usepackage[english]{babel}
\usepackage[utf8]{inputenc}
\usepackage{booktabs}
\usepackage[detect-all=true,group-minimum-digits=4,range-phrase = --,range-units = single]{siunitx} 
\usepackage{amssymb,amsmath,graphicx,chemformula, booktabs, dcolumn}
\usepackage{comment}
\usepackage[pdftex,colorlinks,
allcolors= blue, 
pdftitle={Tunable direct bandgap and optical response in \ch{Mo_{1-x}W_xS2} monolayer alloys: A first-principles investigation}, 
pdfauthor={NDANG AMASSA}]{hyperref} 

\usepackage{cleveref}

\begin{document}
\title{Tunable direct bandgap and optical response in \ch{Mo_{1-x}W_xS2} monolayer alloys: A first-principles investigation}

\author{K. Ndang Amassa}
\email{kevin.ndang@facsciences-uy1.cm}
\affiliation{Department of Physics, Faculty of Science, University of Yaounde I, P.O.Box 812, Yaounde, Cameroon}

\author{J.P. Tchapet Njafa}
\email{jean-pierre.tchapet@facsciences-uy1.cm}
\affiliation{Department of Physics, Faculty of Science, University of Yaounde I, P.O.Box 812, Yaounde, Cameroon}

\author{A. J. Etindele}
\email{anne.etindele@univ-yaounde1.cm}
\affiliation{Higher Teachers Training College, University of Yaounde I, Yaounde P.O. Box 47, Cameroon}

\author{C. Nithaya}
\email{nithaya.chetty@wits.ac.za}
\affiliation{Faculty of Science, University of the Witwatersrand, Johannesburg, South Africa}

\author{S. G. Nana Engo}
\email{serge.nana-engo@facsciences-uy1.cm}
\affiliation{Department of Physics, Faculty of Science, University of Yaounde I, P.O.Box 812, Yaounde, Cameroon}

\date{\today} 

\begin{abstract}
This study presents a comprehensive first-principles investigation of the structural, electronic and optical properties of monolayer \ch{Mo_{1-x}W_xS2} alloys, systematically exploring the full compositional range ($x=0$ to $1$) using density functional theory (DFT). We establish that these alloys are thermodynamically stable and maintain the characteristic 2H crystal structure with minimal structural perturbation upon alloying. A key finding is the preservation of a direct bandgap at the $K$-point across all compositions. This gap exhibits continuous tunability, increasing near-monotonically from \SI{1.696}{\electronvolt} (\ch{MoS2}) to \SI{1.858}{\electronvolt} (\ch{WS2}), a critical feature for tailoring optoelectronic devices. Electronic structure analysis reveals the systematic evolution of the orbital contributions of transition metal $d$ and sulfur $p$ at the edges of the band with composition. Consequently, the optical spectra, evaluated up to \SI{8}{\electronvolt}, show a progressive blueshift in the main features of the interband transition with increasing \ch{W} content, accompanied by predictable changes in key optical constants. Our comprehensive results validate the monolayer \ch{Mo_{1-x} W_xS2} as an electronically versatile platform that offers fine control over electronic and optical properties via alloying, making these tunable direct-gap semiconductors highly promising for next-generation photodetectors, light emitters, and potentially flexible optoelectronic applications exploiting their 2D nature.
\end{abstract}


\maketitle

\section{\label{sec:Intro}Introduction}

Since the isolation of graphene \cite{Novoselov2004,Novoselov2011}, research into two-dimensional (2D) materials has surged, uncovering a diverse landscape rich with unique electronic and optical phenomena. Among the most prominent families in this class are transition metal dichalcogenides (TMDCs) \cite{Wilson1969,Mattheiss1973}. These materials, rapidly evolving from fundamental curiosities to platforms for advanced optoelectronic engineering, have become the focus of intense international investigation, driven by their potential applications spanning electronics, optoelectronics, catalysis, and energy storage \cite{Radisavljevic2011,Butler2013,Duan2024}. TMDCs typically adhere to the chemical formula \ch{MX2}, where a central layer of transition metal atoms (M = Mo, W, Nb, Ta, etc.) is covalently bonded to and sandwiched between two layers of chalcogen atoms (X = S, Se, Te) \cite{Wang2012}. This layered architecture, characterized by relatively weak van der Waals forces between adjacent \ch{MX2} units in the bulk, uniquely allows for exfoliation down to single- or few-layer sheets. In particular, many semiconducting TMDCs, including the archetypal \ch{MoS2} and \ch{WS2}, undergo a fundamental electronic transformation at the monolayer limit, transitioning from an indirect bandgap in their bulk form to a direct bandgap \cite{Wang2012,Chhowalla2013,Duan2024}. This direct-gap nature is particularly advantageous for optoelectronic applications such as photodetectors and light emitters, fueling significant advances in the field \cite{Duan2024}.

Molybdenum disulfide (\ch{MoS2}) has served as a cornerstone for TMDC research. Its monolayer form has a direct bandgap around \SIrange{1.8}{1.9}{\electronvolt}, making it a prime candidate for various electronic and optoelectronic components \cite{Tang2018,Yin2022}. Beyond photonics, \ch{MoS2} has also demonstrated considerable potential in electrocatalysis, particularly for the hydrogen evolution reaction \cite{MoralesGuio2014,Lau2018}, and within energy storage frameworks \cite{Qin2015,Zhou2022}. Despite this promise, pristine \ch{MoS2} faces limitations for practical applications: its inherent bandgap is fixed, restricting tunability for specific device requirements, and its experimentally measured charge carrier mobility often falls short of theoretical predictions, partly due to substrate interactions and material imperfections \cite{Tang2018, Radisavljevic2011}. Although strategies such as alloying have been explored to overcome these limitations \cite{Singh2021}, systematic computational studies of \ch{MoS2}/\ch{WS2} alloys across the full compositional spectrum remain scarce \cite{Chen2023}. This knowledge gap hinders the rational design of compositionally tailored TMDC devices.

Complementing \ch{MoS2}, tungsten disulfide (\ch{WS2})—a close structural relative—presents another compelling monolayer semiconductor featuring a direct bandgap near \SI{2.0}{\electronvolt} \cite{Ganatra2014}. \ch{WS2} exhibits its own suite of promising electronic and optical properties, with recent studies suggesting potential advantages in environmental stability and spin-orbit coupling effects compared to \ch{MoS2} \cite{Rozhansky2024}. The inherent similarities between \ch{MoS2} and \ch{WS2}, coupled with their distinct individual characteristics, make their combination by alloying a highly attractive strategy for creating novel materials with potentially synergistic or precisely tunable functionalities \cite{Komsa2013,Duan2024}. Although vertical heterostructures of \ch{MoS2}/\ch{WS2} have been extensively studied \cite{Ma2021}, substitutional alloying within a single monolayer, where \ch{Mo} and \ch{W} atoms randomly occupy lattice sites, offers distinct advantages. This approach enables homogeneous property tuning throughout the material and eliminates the interlayer transport barriers inherent in heterostructures \cite{Liu2024}, eliminating the need for precise interfacial control. This alloying strategy, central to our work, remains comparatively underexplored in computational studies despite the demonstrated experimental feasibility \cite{Liu2024}.

Alloying within the TMDC family, by adjusting the compositional ratio $x$ \cite{Singh2021}, has emerged as a powerful method for continuously tuning electronic and optical properties. Such alloys, including the \ch{MoS2}/\ch{WS2} system, have demonstrated good thermodynamic stability \cite{Liu2017}, with recent advances in chemical vapor deposition enabling precise control over stoichiometry \cite{Liu2024}. The ability to engineer the bandgap across the visible spectrum by varying the composition is a key motivator, promising significant improvements for wavelength-tunable photodetectors and customized light sources \cite{Lin2023}. However, a significant limitation in previous theoretical work is the frequent focus on only a few discrete compositions, such as the endpoints ($x=0, 1$) and the midpoint ($x=0.5$) (see, e.g., Ref. \cite{Chen2023}). This approach leaves the continuous evolution of properties across the full alloy series poorly understood, hindering the development of predictive models for rational device design. To our knowledge, this work represents the first comprehensive DFT study to systematically map the structural, electronic, and optical properties throughout the \textit{complete} compositional range ($x=0$ to $1$) of monolayer \ch{Mo_{1-x}W_xS2} alloys, thereby addressing this critical knowledge gap. We employ first-principles density functional theory (DFT) to conduct this investigation, focusing on three critical questions: (i) How does the substitution of \ch{W} affect the lattice parameters and stability of the monolayer \ch{MoS2}? (ii) Can the direct bandgap be linearly tuned across the \ch{MoS2}–\ch{WS2} spectrum? (iii) How do alloy-driven electronic changes manifest in optical properties such as absorption and dielectric response? By elucidating how progressive substitution of \ch{Mo} with \ch{W} atoms influences lattice structure, electronic band alignment, and optical response, this work provides a roadmap for targeted property engineering in 2D semiconductors. The findings bear significant implications for the development of flexible nanoelectronics and tunable optoelectronic devices \cite{Aftab2023}.

The remainder of this paper is structured as follows. \Cref{sec:Methods} outlines the theoretical and computational methods employed, including details of the DFT calculations performed using Quantum ESPRESSO and the supercell construction methodology. \Cref{sec:Results} presents and discusses the calculated structural, electronic, and optical trends observed throughout the series of alloys. \Cref{sec:Conclusion} summarizes the key findings and discusses their implications for future applications in flexible optoelectronics and advanced heterostructure design.
 
\section{\label{sec:Methods}Theoretical methods}

First-principles calculations were performed using the plane wave self-consistent field (PWscf) code within the Quantum ESPRESSO software suite (version 7.3) \cite{Giannozzi2009, Giannozzi2017}, which is based on density functional theory (DFT). The Perdew-Burke-Ernzerhof (PBE) parameterization of the generalized gradient approximation (GGA) was used for the exchange-correlation functional \cite{Perdew1996}. To account for van der Waals interactions, empirical dispersion corrections DFT-D3 proposed by Grimme \cite{Grimme2006} were included. This inclusion ensures comparability with previous studies of TMDCs and provides a consistent framework for future investigations. Furthermore, the inclusion of dispersion corrections does not adversely affect monolayer calculations and provides a unified framework for potential future investigations involving interfaces or substrate interactions.

Norm-conserving pseudopotentials from PSLibrary \cite{Corso2014} were utilized, treating the following valence electrons: \ch{Mo} ($4p^64d^5 5s^1$), \ch{W} ($5p^6 5d^4 6s^2$), and \ch{S} ($3s^2 3p^4$). Specifically, the Projector Augmented Wave (PAW) pseudopotential was employed for geometry optimization and the calculation of electronic properties, while the Optimized Norm-Conserving Vanderbilt (ONCV) pseudopotential was preferred for the calculation of optical properties, leveraging the strengths of each type for the respective calculations. Cross-checks confirmed that the choice of pseudopotential did not significantly alter the fundamental electronic structure features relevant to this study. The electronic wave functions were expanded using a set of plane wave bases with a kinetic energy cut-off of \SI{70}{Ry}, determined through convergence tests. Integration in the Brillouin zone was performed using a Monkhorst-Pack \cite{Monkhorst1976} mesh of $6 \times 6 \times 1$ $K$-points centered at the $\Gamma$-point for supercell calculations; this mesh was verified to ensure energy convergence to within \SI{0.01}{\electronvolt} per atom. Structural optimizations were performed until the total energy converged to within \SI{1e-7}{\electronvolt} per supercell and the residual forces on all atoms were below \SI{1e-3}{eV/\angstrom}, thresholds consistent with high-accuracy DFT studies of 2D materials \cite{Liu2024}.

The pristine monolayer structures of hexagonal molybdenum disulfide (\ch{MoS2}) and tungsten disulfide (\ch{WS2}) served as the basis for our models. We used optimized lattice parameters of $a = b = \SI{3.18}{\angstrom}$ for both monolayers, consistent with typical DFT-PBE calculations \cite{Komsa2013}. To simulate isolated monolayers and prevent spurious interactions between periodic images, a vacuum spacing of $c = \SI{20}{\angstrom}$ was introduced perpendicular to the atomic plane. These optimized monolayer parameters were then used to construct $4\times4\times1$ supercells to model the \ch{Mo_{1-x}W_xS2} alloys. Each supercell contained a total of 48 atoms: 16 transition metal sites (occupied by \ch{Mo} or \ch{W}) and 32 sulfur (S) atoms. We investigated various compositions by substituting \ch{Mo} atoms with \ch{W} atoms at the metal sites. Specifically, configurations corresponding to \ch{W} concentrations of \SIlist{6.25; 12.5; 25; 37.5; 50; 62.5; 75; 87.5}{\percent} were generated by placing \ch{W} atoms at representative substitution sites within the $4\times4$ metal lattice. The atomic positions and lattice vectors were fully relaxed for each configuration.

\begin{figure*}[htpb]
	\centering
	\includegraphics[width=.9\linewidth]{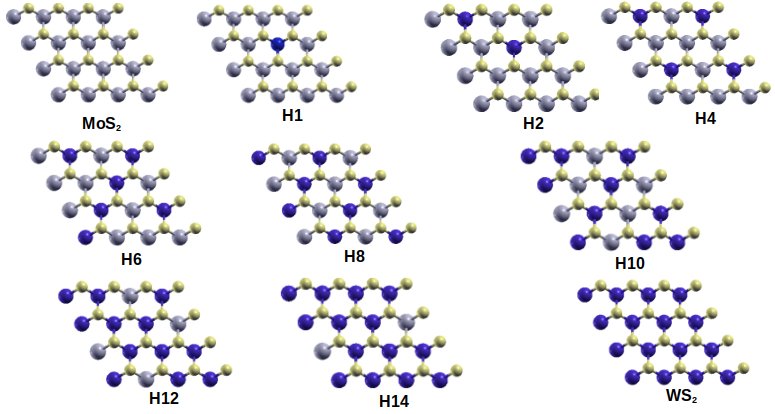}	
    \caption{(Colour online). Top view schematics of the $4\times4\times1$ supercell configurations used to model \ch{Mo_{1-x}W_xS2} alloys at different \ch{W} concentrations ($x$). Yellow, gray, and blue spheres represent \ch{S}, \ch{Mo}, and \ch{W} atoms, respectively. Configurations labeled H1 ($x = \num{0.0625}$), H2 ($x = \num{0.125}$), H4 ($x =\num{0.25}$), H6 ($x =\num{0.375}$), H8 ($x =\num{0.5}$), H10 ($x =\num{0.625}$), H12 ($x =\num{0.75}$), and H14 ($x =\num{0.875}$) illustrate the progressive substitution of \ch{Mo} by \ch{W}.}
	\label{fig:Structure}
\end{figure*}

The specific atomic arrangements for each composition studied are depicted in \Cref{fig:Structure}. It is important to acknowledge the limitations inherent in modeling random alloys with finite supercells. Our use of a single specific atomic configuration for each composition, chosen to promote homogeneity, is a common necessity due to computational cost but does not fully capture the ensemble average over all possible random arrangements. Consequently, the calculated properties might exhibit some dependence on the specific configuration chosen, particularly for properties sensitive to local environments. Future work employing techniques such as special quasirandom structures (SQS) \cite{Zunger1990}, designed to mimic the local correlations of a random alloy, or averaging multiple distinct configurations for each composition could provide a more rigorous treatment of configurational disorder, although this falls beyond the scope of the present study.

Following optimization of the geometry of each alloy structure using the DFT settings described above, the electronic band structures and optical properties were subsequently calculated.

\subsection{Optical properties calculations}

The optical properties were derived from the complex dielectric function dependent on the frequency, $\varepsilon(\omega) = \varepsilon_1(\omega) + i\varepsilon_2(\omega)$, calculated within the random phase approximation (RPA) as implemented in the \textit{epsilon.x} code of Quantum ESPRESSO, using the electronic structure obtained from the ONCV pseudopotentials. The imaginary part, $\varepsilon_2(\omega)$, is determined directly from transitions between occupied ($\psi_\alpha$, energy $E_\alpha$) and unoccupied ($\psi_\beta$, energy $E_\beta$) Kohn-Sham states in the long-wavelength limit ($\mathbf{q} \to 0$):
\begin{equation}\label{eq:eq1} 
\begin{split}
    \varepsilon_2(\omega) &= \frac{2e^2\pi}{\varepsilon_0 m^2 \omega^2 V} \sum_{\alpha,\beta} |\langle\psi_\beta|\vec{e}\cdot\vec{p}|\psi_\alpha\rangle|^2\\
    &\times [f(E_\alpha) - f(E_\beta)] \delta(E_\beta - E_\alpha - \hbar\omega),\\
\end{split}
\end{equation}
where $e$ is the elementary charge, $\varepsilon_0$ is the vacuum permittivity, $m$ is the electron mass, $V$ is the volume of the computational cell, $\vec{e}$ is the polarization vector of the incident electric field (aligned in-plane for monolayer TMDCs), $\vec{p}$ is the momentum operator, $\hbar\omega$ is the photon energy, and $f(E)$ is the Fermi-Dirac distribution function. The Dirac delta function is approximated by a Gaussian smearing of width $\eta = \SI{0.1}{\electronvolt}$ to account for finite-lifetime effects. The real part, $\varepsilon_1(\omega)$, is then obtained via the Kramers-Kronig relation:
\begin{equation}\label{eq:kk}
\varepsilon_1(\omega) = 1 + \frac2\pi \mathcal{P} \int_0^\infty \frac{\omega' \varepsilon_2(\omega')}{\omega'^2 - \omega^2} d\omega',
\end{equation}
where $\mathcal{P}$ denotes the Cauchy principal value of the integral.

From the complex dielectric function, the complex refractive index $\tilde{n} = n + ik$ is determined using the standard relations:
\begin{align}\label{eq:nk_from_eps}
&n^2 = \frac12\left(|\varepsilon| + \varepsilon_1\right),
&& k^2 = \frac12\left(|\varepsilon| - \varepsilon_1\right),
\end{align}
where $|\varepsilon|^2 = \varepsilon_1^2 + \varepsilon_2^2$. $n$ is the real refractive index, and $k$ is the extinction coefficient. For analysis under normal incidence conditions, the absorption coefficient ($\alpha$), reflectance ($R$), transmittance ($T$), and optical conductivity ($\sigma$) are calculated using:
\begin{align}
&\alpha(\omega) = \frac{2\omega k(\omega)}{c_0} = \frac{4\pi k(\lambda)}{\lambda}, \label{eq:alpha_formula}\\
& R(\omega) = \frac{[n(\omega) - 1]^2 + k(\omega)^2}{[n(\omega) + 1]^2 + k(\omega)^2}, \label{eq:reflectance_formula}\\ 
& T(\omega) = [1 - R(\omega)]e^{-\alpha(\omega)t}, \label{eq:transmittance_formula}\\ 
& \sigma(\omega) = \frac{n(\omega)\alpha(\omega)c_0 \varepsilon_0}{2}, \label{eq:conductivity_formula_v2} 
\end{align}
where $t = \SI{0.65}{\nm}$ is the monolayer thickness \cite{Wang2012}, $\lambda = 2\pi c_0/\omega$ is the vacuum wavelength, and $c_0$ is the speed of light.

Importantly, it must be emphasized that standard DFT calculations coupled with the RPA formalism for optical properties do not capture excitonic effects, which arise from the Coulomb interaction between the excited electron and the remaining hole. These effects are known to be particularly strong in 2D TMDCs, leading to significant exciton binding energies (typically \SIrange{0.3}{0.5}{\electronvolt} or more) \cite{Chernikov2014, Qiu2013, Komsa2013}. Consequently, the calculated optical absorption peaks presented in this work (derived from interband transitions within RPA) will appear at energies higher than those observed experimentally in photoluminescence or absorption spectra. Although our calculations accurately capture trends in interband transitions due to alloying, a direct comparison of absolute peak energies with experimental excitonic peaks requires higher-level theoretical methods like the GW approximation combined with the Bethe-Salpeter equation (GW-BSE) \cite{Komsa2013}, which explicitly include electron-hole interactions. The discussion of the optical results in \Cref{sec:Results} should be interpreted with this crucial limitation in mind.

\section{\label{sec:Results}Results and discussion} 

Having established the computational methodology in \Cref{sec:Methods}, we now turn to the central results of our study on \ch{Mo_{1-x}W_xS2} monolayer alloys. This section explores the predicted structural, electronic, and optical properties, with the aim of elucidating the consequences of compositional tuning in this important TMDC alloy system. We commence our discussion by evaluating the structural integrity and energetic stability of the alloy configurations. Subsequently, we investigate the electronic band structures and the density of states to understand how alloying modifies the fundamental electronic character, particularly the critical bandgap properties. Building upon this electronic foundation, we then analyze the optical response, examining how the absorption and dielectric properties evolve with varying tungsten concentration.

\subsection{\label{subsec:Structural}Structural analysis}

Following the computational methodology detailed in \Cref{sec:Methods}, we first investigated the structural properties of the pristine \ch{MoS2} and \ch{WS2} monolayers alongside their substitutional alloys, \ch{Mo_{1-x}W_xS2}, across the full composition range ($x$ from 0 to 1). The specific atomic configurations used for alloy supercells (denoted H1, H2, H4, \ldots, H14, corresponding to increasing \ch{W} concentrations, where H$k$ signifies $k$ \ch{W} atoms per 16 metal sites) are visualized in \Cref{fig:Structure}.

Modeling the properties of random alloys presents inherent computational challenges. An ideal approach would involve averaging over a large ensemble of possible atomic configurations for each composition within the supercell, but this is often computationally prohibitive. Therefore, following a common strategy, we selected a single representative atomic configuration (H1 to H14, as shown in \Cref{fig:Structure}) for each concentration studied. These specific configurations were chosen to ensure a reasonably homogeneous distribution of the substituted tungsten atoms within the $4 \times 4$ supercell lattice (16 metal sites), thereby minimizing artificial effects from clustering or excessive ordering. This approach aims to capture properties representative of the intended disordered alloy state, while acknowledging the inherent limitations imposed by using a periodic supercell of finite size. Although an exhaustive search for the absolute ground-state configuration was not performed for each stoichiometry, the selected structures represent plausible and relatively dispersed atomic arrangements.

In all optimized models, transition metal atoms (\ch{Mo}/\ch{W}) form a hexagonal lattice sandwiched between two layers of sulfur (S) atoms, consistently maintaining the characteristic 2H phase structure common to both parent compounds. It is important to reiterate that this substitutional alloying involves isoelectronic \ch{Mo} and \ch{W} atoms (both in Group 6), which means that it does not introduce additional charge carriers (electrons or holes) into the system, allowing a focused study on the effects of atomic substitution itself.

The key structural parameters obtained following the optimization of the geometry, summarized in \Cref{tab:StructuralParams}, reveal a remarkable structural consistency throughout the series of alloys. Individual Mo-S and W-S bond lengths remain stable within very narrow ranges (Mo-S: \SIrange{2.412}{2.416}{\angstrom}, W-S: \SIrange{2.410}{2.418}{\angstrom}), closely matching their respective values in pristine binary monolayers. This stability indicates that the local coordination geometry around both the Mo and W atoms is largely preserved upon alloying. Likewise, the average transition metal-metal distances (Mo-Mo, W-W, Mo-W), which dictate the effective in-plane lattice parameter 'a', exhibit only minimal variation, remaining centered around \SI{3.18}{\angstrom} throughout the entire composition range. While a slight lattice expansion might be anticipated when substituting the nominally smaller Mo atom with the larger W atom, this effect is subtle in the fully relaxed structures, suggesting that the S sublattice effectively accommodates the substitution with minimal strain. The S-S distance, reflecting the thickness of the monolayer, also shows only marginal changes, increasing slightly from \SI{3.127}{\angstrom} ($x=0$) to \SI{3.147}{\angstrom} ($x=1$) with increasing \ch{W} content.

These calculated structural parameters compare favorably with the available experimental data, lending further confidence to our models. For instance, our calculated Mo-S bond length near \SI{2.41}{\angstrom} aligns closely with experimental measurements (\SI{2.41(2)}{\angstrom}) derived from scanning transmission electron microscopy (STEM) studies \cite{Sohn2023}. Furthermore, our prediction of minimal structural perturbation upon alloying resonates with experimental observations in related alloy systems, such as \ch{Mo_{1-x}W_xSe2} \cite{Wells2024,Bai2024}.

To quantitatively assess the thermodynamic stability of these alloys, we calculated the formation energy ($E_b$) per atom relative to the isolated constituent atoms using the following expression:
\begin{equation}\label{eq:FormationEnergy}
E_b = \frac1N \left[ E_H - n_S E_{\ch{S}} - n_{Mo} E_{\ch{Mo}} - n_W E_{\ch{W}} \right].
\end{equation}
Here, $E_H$ is the total energy of the hybrid alloy supercell, $E_{\ch{S}}$, $E_{\ch{Mo}}$, and $E_{\ch{W}}$ represent the calculated total energies of isolated sulfur, molybdenum and tungsten atoms, respectively. The terms $n_S$, $n_{Mo}$, and $n_W$ denote the number of sulfur atoms (32), molybdenum atoms, and tungsten atoms within the supercell ($n_{Mo} + n_W = 16$), and $N = n_S + n_{Mo} + n_W$ is the total number of atoms in the supercell (48).

The calculated formation energies per atom, listed in the final column of \Cref{tab:StructuralParams}, provide critical insight into the thermodynamic stability. Importantly, $E_b$ is negative for all alloy compositions investigated, confirming that alloy formation is energetically favorable compared to the constituent isolated atoms. The values range from \SI{-6.910}{eV/atom} (as per Table I value for MoS2) for pure \ch{MoS2} to \SI{-7.370}{eV/atom} (as per Table I value for WS2, previously -7.24, then -7.399) for pure \ch{WS2}. Interestingly, the formation energy generally becomes more negative (indicating enhanced stability) as \ch{W} content increases, reaching a minimum value of \SI{-7.280}{eV/atom} (as per Table I) near the W-rich end ($x=0.875$, configuration H14). This suggests a slight energetic preference for incorporating \ch{W} into the \ch{MoS2} lattice, although the overall variation in stability across the entire composition range is modest (within \SI{\sim 0.46}{eV/atom} comparing MoS2 to WS2 based on Table I, or \SI{\sim 0.37}{eV/atom} comparing MoS2 to H14). Taken together, the structural optimizations showing the preservation of the lattice parameters and the consistently negative formation energies firmly establish that these \ch{Mo_{1-x}W_xS2} alloys maintain the characteristic 2H structure and are thermodynamically stable throughout the full compositional range, making them viable for experimental synthesis and device integration.

\begin{table*}[htpb]
\centering
\caption{Calculated structural parameters and formation energies for \ch{Mo_{1-x}W_xS2} alloys. Bond lengths (Mo-Mo, Mo-W, W-W, Mo-S, W-S) and the S-S distance are averaged values in \unit{\angstrom}. The formation energy $E_b$ (Eq. \eqref{eq:FormationEnergy}) is given in \unit{eV/atom}.}
\label{tab:StructuralParams}
\sisetup{round-mode=places, round-precision=3}
\begin{tabular}{@{}lccccccc@{}}
\toprule
Structure & Mo-Mo & Mo-W & W-W & Mo-S & W-S & S-S & $E_b$ \\
 & {(\unit{\angstrom})} & {(\unit{\angstrom})} & {(\unit{\angstrom})} & {(\unit{\angstrom})} & {(\unit{\angstrom})} & {(\unit{\angstrom})} & {(\unit{eV/atom})} \\
\midrule
\ch{MoS2} ($x=0$)     & \num{3.180} & --          & --          & \num{2.412} & --          & \num{3.129} & \num{-6.910} \\
H1 (\SI{6.25}{\percent} W) & \num{3.184} & \num{3.184} & --          & \num{2.416} & \num{2.412} & \num{3.127} & \num{-6.970} \\
H2 (\SI{12.5}{\percent} W) & \num{3.182} & \num{3.183} & --          & \num{2.416} & \num{2.410} & \num{3.128} & \num{-6.980} \\
H4 (\SI{25.0}{\percent} W) & \num{3.180} & \num{3.180} & --          & \num{2.415} & \num{2.411} & \num{3.129} & \num{-7.040} \\
H6 (\SI{37.5}{\percent} W) & \num{3.185} & \num{3.182} & \num{3.180} & \num{2.413} & \num{2.413} & \num{3.132} & \num{-7.070} \\
H8 (\SI{50.0}{\percent} W) & \num{3.180} & \num{3.180} & \num{3.180} & \num{2.415} & \num{2.415} & \num{3.133} & \num{-7.130} \\
H10 (\SI{62.5}{\percent} W)& \num{3.176} & \num{3.183} & \num{3.168} & \num{2.416} & \num{2.413} & \num{3.132} & \num{-7.180} \\
H12 (\SI{75.0}{\percent} W)& --          & \num{3.179} & \num{3.184} & \num{2.415} & \num{2.413} & \num{3.134} & \num{-7.240} \\
H14 (\SI{87.5}{\percent} W)& --          & \num{3.177} & \num{3.178} & \num{2.416} & \num{2.413} & \num{3.136} & \num{-7.280} \\
\ch{WS2} ($x=1$)      & --          & --          & \num{3.180} & --          & \num{2.418} & \num{3.147} & \num{-7.370} \\ 
\bottomrule
\end{tabular}
\end{table*}

\subsection{\label{subsec:Electronic}Electronic structure}

\begin{figure*}[htpb]
	\centering
    \includegraphics[width=.9\linewidth]{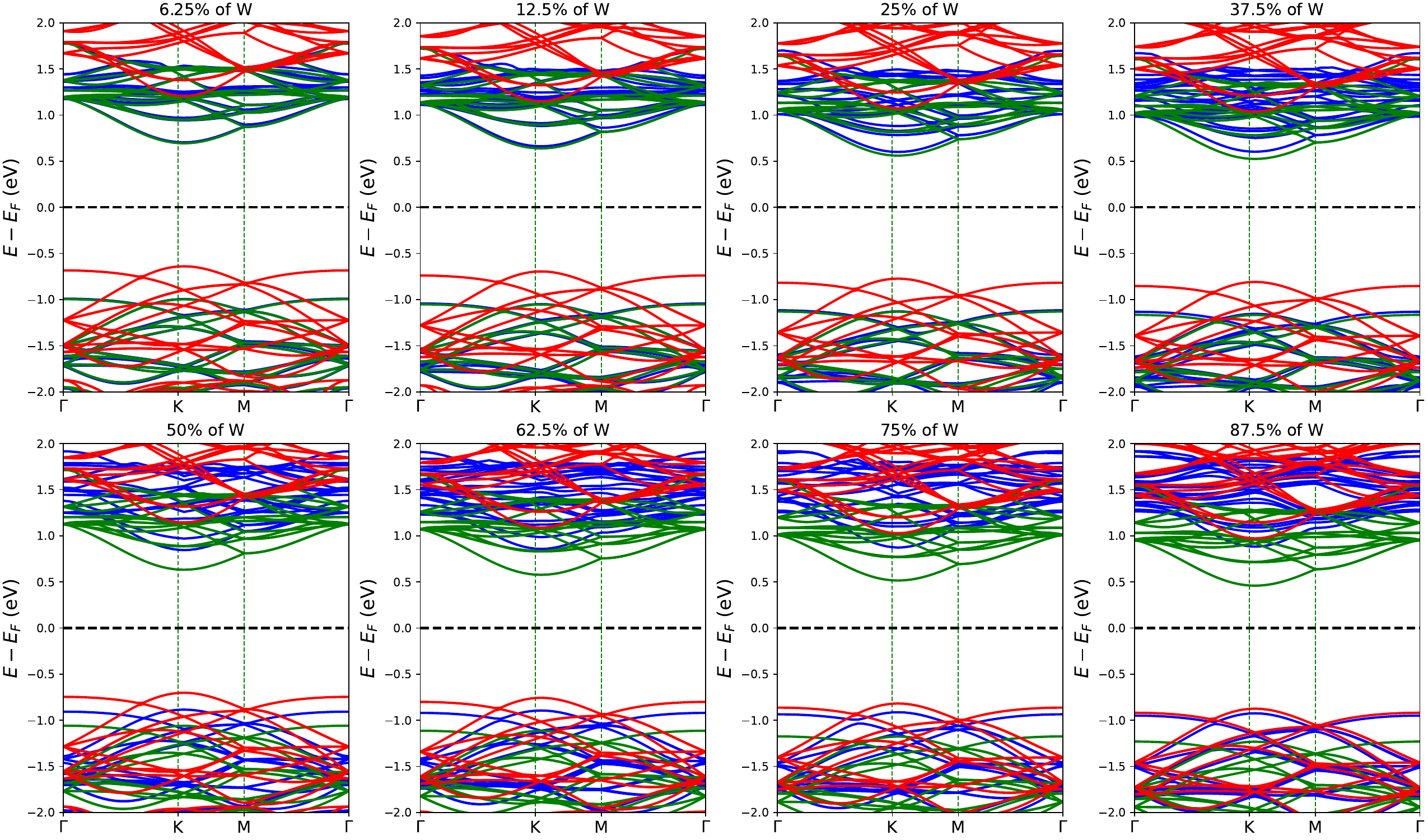} 
	\caption{(Colour online). Calculated electronic band structures of \ch{Mo_{1-x}W_xS2} monolayer alloys for different \ch{W} concentrations. The Fermi level (dashed lines) is set to \SI{0}{\electronvolt}. High symmetry points $\Gamma$, K, M define the path in the Brillouin zone. The direct bandgap at the $K$-point is evident for all compositions.}
    \label{fig:Band structure} 
\end{figure*}

\begin{figure*}[htpb]
	\centering
    \includegraphics[width=1.\linewidth]{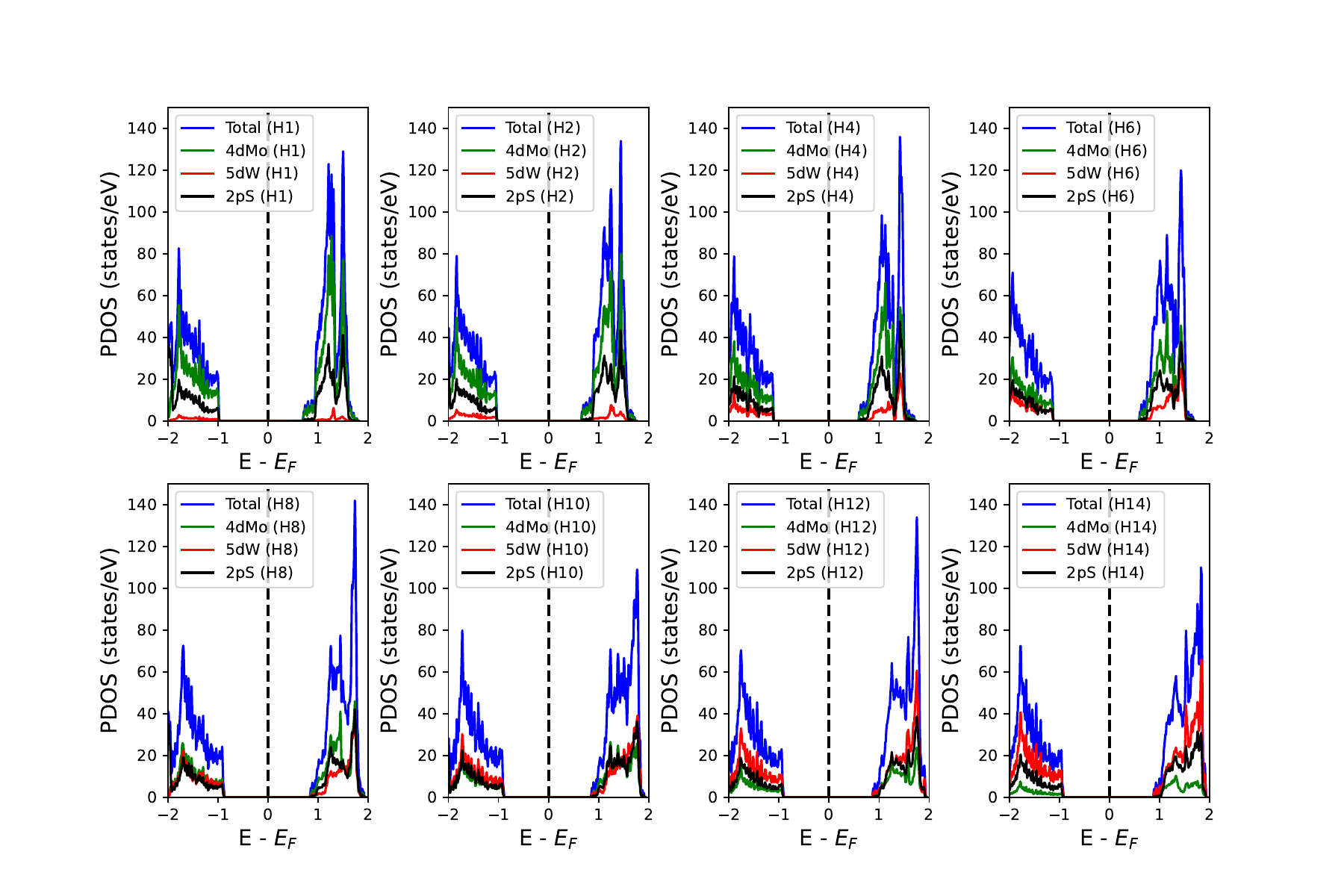}
	\caption{(Colour online). Projected Density of States (PDOS) for \ch{Mo_{1-x}W_xS2} alloys. Contributions from \ch{Mo} d-states (e.g., blue), \ch{W} d-states (e.g., red), and S p-states (e.g., green) are shown. The Fermi level (dashed lines) is set to \SI{0}{\electronvolt}. The plots illustrate the evolution of electronic states near the band edges with changing \ch{W} concentration.}
    \label{fig:DOS} 
\end{figure*}

To elucidate the impact of alloying on electronic characteristics, we calculated the electronic band structures and projected density of states (PDOS) for pristine monolayers and various \ch{Mo_{1-x}W_xS2} alloy compositions. The calculated band structures are presented in \Cref{fig:Band structure}, and the corresponding PDOS are shown in \Cref{fig:DOS}.

A key electronic feature revealed by band structure calculations (\Cref{fig:Band structure}) is that all alloy configurations retain a direct bandgap character. For all compositions studied, the conduction band minimum (CBM) and the valence band maximum (VBM) are located at the high-symmetry $K$-point (or points equivalent by symmetry) within the Brillouin zone. This preservation of the direct gap nature across the alloy series is highly significant for optoelectronic applications.

Furthermore, the magnitude of the direct bandgap exhibits a clear and systematic dependence on the alloy composition. The calculated direct Kohn-Sham bandgaps at the $K$-point are found to be \SI{1.696}{\electronvolt} (pure \ch{MoS2}), \SI{1.700}{\electronvolt} (H1, \SI{6.25}{\percent} W),\SI{1.709}{\electronvolt} (H2, \SI{12.5}{\percent} W), \SI{1.723}{\electronvolt} (H4, \SI{25}{\percent} W), \SI{1.741}{\electronvolt} (H6, \SI{37.5}{\percent} W), \SI{1.737}{\electronvolt} (H8, \SI{50.0}{\percent} W), \SI{1.760}{\electronvolt} (H10, \SI{62.5}{\percent} W), \SI{1.794}{\electronvolt} (H12, \SI{75}{\percent} W),\SI{1.818}{\electronvolt} (H14, \SI{87.5}{\percent} W), and \SI{1.858}{\electronvolt} (pure \ch{WS2}). These values demonstrate a near-monotonic increase in the bandgap energy as the concentration of tungsten (W) atoms increases. This trend reflects the gradual transition of the electronic structure, reflecting the progressive substitution of \ch{Mo} with W, since the pristine \ch{WS2} possesses a slightly larger intrinsic bandgap than \ch{MoS2}.

\begin{figure*}[htpb]
	\centering
    \includegraphics[width=.6\linewidth]{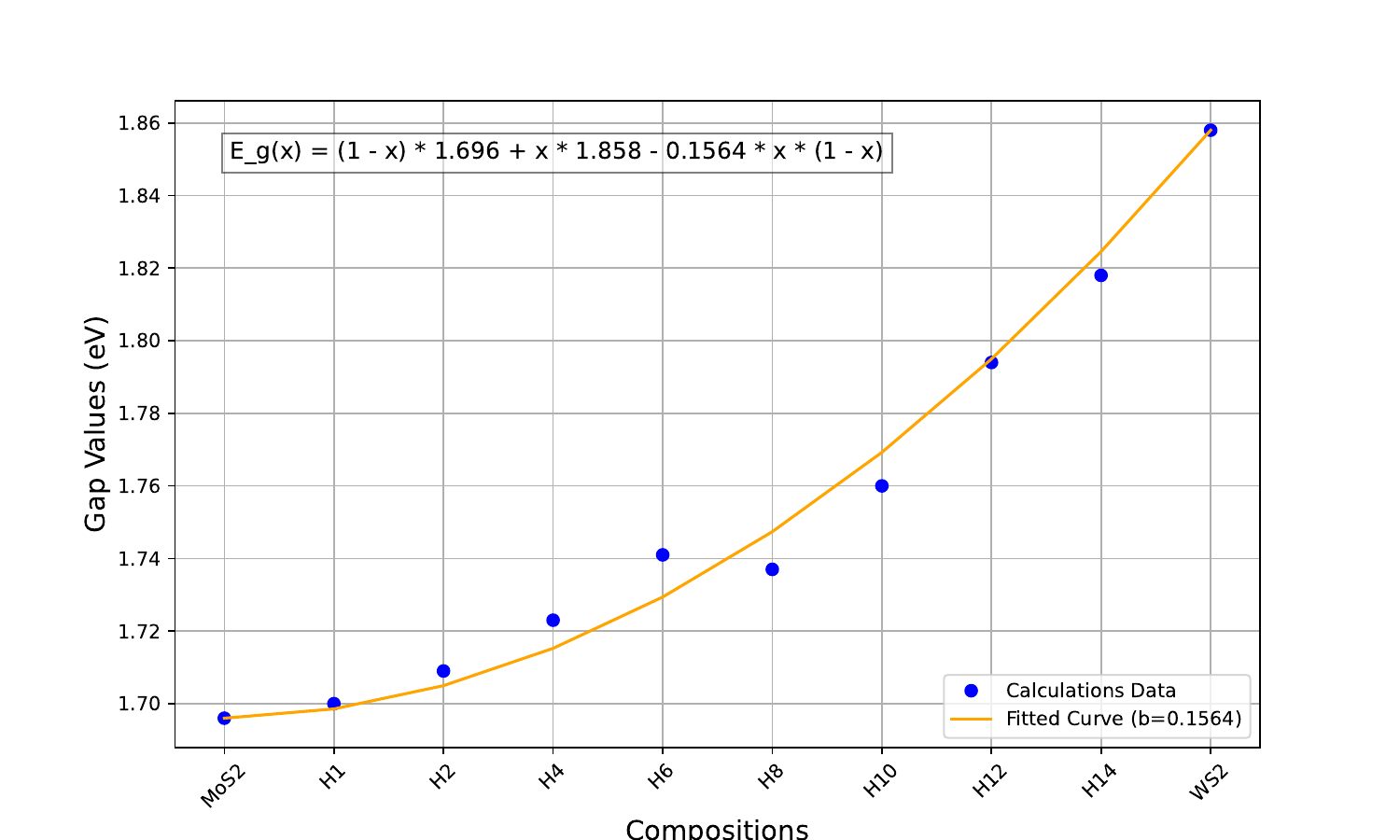} 
	\caption{(Colour online). Evolution of the band gap energy $E_g$ as a function of composition for various structures, including \ch{MoS2}, H1, H2, H4, H6, H8, H10, H12, H14, and \ch{WS2}. The fitted curve represents the relationship 
$E_g(x)=(1-x)\num{1.696} + x\num{1.858} -\num{0.1564}x(1-x)$, illustrating how the band gap varies with composition. The values of the band gap range from \SI{1.696}{\electronvolt} to \SI{1.858}{\electronvolt} across the different materials.}
    \label{fig:gap_evolution} 
\end{figure*}

Although the increase appears largely linear, closer inspection reveals a slight deviation from perfect linearity, exhibiting a phenomenon known as bandgap bowing. This is common in semiconductor alloys where properties do not strictly follow the linear interpolation predicted by the virtual crystal approximation (VCA) or Vegard's law applied to the bandgap. This bowing typically arises from factors such as local structural relaxations around substituted atoms, differences in electronegativity that lead to charge transfer effects, and intrinsic electronic disorder within the alloy \cite{Popescu2010}. Comparing the calculated gap at $x=0.5$ (H8, \SI{1.737}{\electronvolt}) with the value obtained from the linear interpolation between \ch{MoS2} and \ch{WS2} (\SI{1.777}{\electronvolt}) suggests a small downward bowing (i.e., the actual gap is slightly lower than the linear average). This nonlinearity can often be described by adding a quadratic term, $E_g(x) = (1-x)E_g(0) + xE_g(1) - bx(1-x)$, where $b$ is the bowing parameter. Although the bowing effect observed here is relatively small compared to some III-V alloys, acknowledging its presence is important for precise bandgap engineering based on these materials. The evolution of the band gap energy as a function of composition for various structures, including \ch{MoS2}, H1, H2, H4, H6, H8, H10, H12, H14, and \ch{WS2} are shown in \Cref{fig:gap_evolution}, where the corresponding bowing parameter is $b = 0.1564$.

It is important to recognize that these calculated DFT-PBE gaps (\SIrange{1.696}{1.858}{\electronvolt}) represent the Kohn-Sham gap and, as expected, systematically underestimate the known experimental optical gaps (typically \SIrange{1.85}{2.10}{\electronvolt} for \ch{MoS2}/\ch{WS2} monolayers \cite{Mak2010, Chernikov2014}), a common limitation of the PBE functional for predicting excitation energies. Specifically, the calculated bandgap for pristine \ch{MoS2} (\SI{1.696}{\electronvolt}) underestimates representative experimental optical gap values (\SI{\sim 1.9}{\electronvolt}) \cite{Splendiani2010} by approximately \SI{0.2}{\electronvolt}, consistent with typical DFT-PBE limitations for TMDCs. However, the key aspect is the predicted composition-dependent trend. Our calculations predict a total tuning range of approximately \SI{0.162}{\electronvolt} (using precise end values) across the series ($E_g(x=1) - E_g(x=0)$). This value aligns remarkably well with the tuning range observed experimentally via photoluminescence (PL) measurements on CVD-grown \ch{Mo_{1-x}W_xS2} alloys \cite{Liu2017}. This quantitative agreement in the relative tuning effect provides strong validation that the DFT-PBE approach, despite its limitations in predicting absolute gap values, reliably captures the compositional engineering potential essential for device design.

\begin{table}[htbp] 
    \centering
    \caption{Comparison of calculated DFT-PBE Kohn-Sham bandgaps ($E_g^{\text{PBE}}$) with representative experimental optical gaps ($E_g^{\text{Opt, Exp}}$) and calculated GW fundamental gaps ($E_g^{\text{GW}}$) for pristine \ch{MoS2} and \ch{WS2} monolayers.}
    \label{tab:GapComparison}
    \sisetup{round-mode=places, round-precision=3}
    \begin{tabular}{@{}lccc@{}}
        \toprule
        Material & $E_g^{\text{PBE}}$ (\unit{\electronvolt}) & $E_g^{\text{Opt, Exp}}$ (\unit{\electronvolt}) & $E_g^{\text{GW}}$ (\unit{\electronvolt}) \\
        \midrule
        \ch{MoS2} & \num{1.696} & \numrange{1.85}{1.90} \cite{Mak2010, Splendiani2010} & \num{\sim 2.8} \cite{Qiu2013, Komsa2013} \\
        \ch{WS2}  & \num{1.858} & \numrange{1.95}{2.10} \cite{Chernikov2014, Zhao2013} & \num{\sim 2.7} \cite{Qiu2013, Komsa2013} \\
        \bottomrule
    \end{tabular}
\end{table}

This controllable adjustment of the direct bandgap via composition is, therefore, a significant advantage offered by these alloyed TMDCs \cite{Popescu2010, Liu2017}. The PDOS plots (\Cref{fig:DOS}) clearly illustrate the origins of the band edges and their evolution with composition. The region of the valence band maximum (VBM) is mainly composed of hybridized transition metal $d$ orbitals (\ch{Mo} $4d$ and \ch{W} $5d$) and sulfur $p$ orbitals. As the tungsten concentration ($x$) increases, the contribution of \ch{W} $5d$ states to the VBM increases progressively, while the contribution of \ch{Mo} $4d$ states diminishes. Similarly, the conduction band minimum (CBM) region is dominated by transition metal $d$ orbitals. Here also, increasing $x$ leads to a greater contribution from the \ch{W} $5d$ states and a lesser contribution from the \ch{Mo} $4d$ states at the edge of the CBM. The underlying physical origin for the increase in the bandgap with \ch{W} content can be attributed, in part, to relativistic effects and potentially deeper energy levels of the \ch{W} $5d$ orbitals compared to the \ch{Mo} $4d$ orbitals, particularly influencing the position of the conduction band minimum, as supported by the PDOS analysis (\Cref{fig:DOS}) showing shifts in the relevant $d$-state contributions.

\subsection{\label{subsec:Optical}Optical response}

\begin{figure*}[htpb]
	\centering
    \includegraphics[width=.9\linewidth]{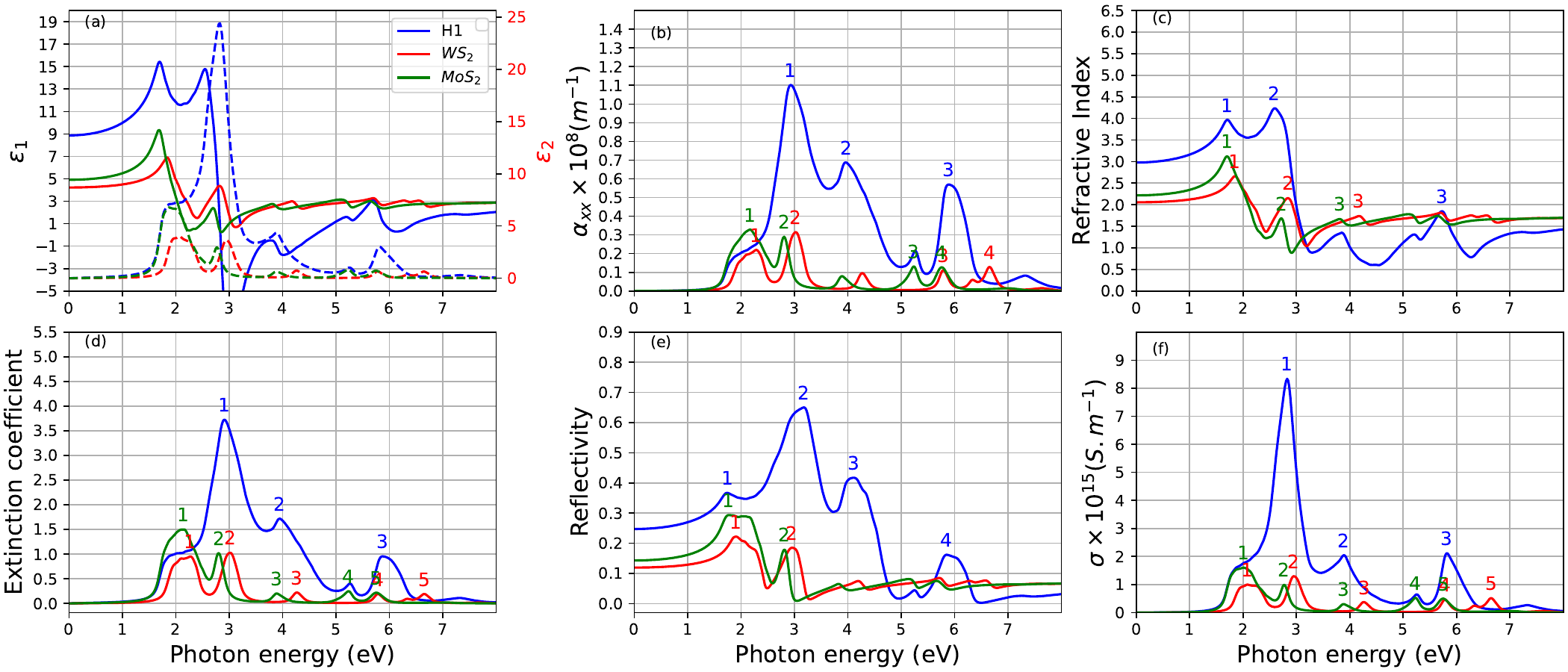}
	\caption{(Colour online). Calculated optical properties for the H1 alloy (\ch{Mo_{0.9375}W_{0.0625}S2}, \SI{6.25}{\percent} W). (a) Real ($\epsilon_1$, solid line) and imaginary ($\epsilon_2$, dashed line) parts of the dielectric function vs. photon energy. Key features in $\epsilon_2$ include peaks corresponding to the A and B transitions at approx. \SI{2.14}{\electronvolt} and \SI{2.36}{\electronvolt}, respectively. (b) Absorption coefficient ($\alpha$) vs. photon energy, showing significant absorption onset near \SI{1.8}{\electronvolt} and peaks corresponding to A and B transitions. (c) Real part of the refractive index ($n$) vs. photon energy, exhibiting prominent dispersive features around the A/B transition energies. (d) Extinction coefficient ($k$) vs. photon energy, mirroring the absorption spectrum structure with peaks at approx. \SI{2.14}{\electronvolt} and \SI{2.36}{\electronvolt}. (e) Normal incidence Reflectivity ($R$) vs. photon energy, showing peaks related to high absorption/refractive index regions. (f) Optical conductivity ($\sigma$) vs. photon energy, displaying peaks corresponding to the main absorption features.}
    \label{fig:h1-optic} 
\end{figure*}

\begin{figure*}[htpb]
	\centering
    \includegraphics[width=.9\linewidth]{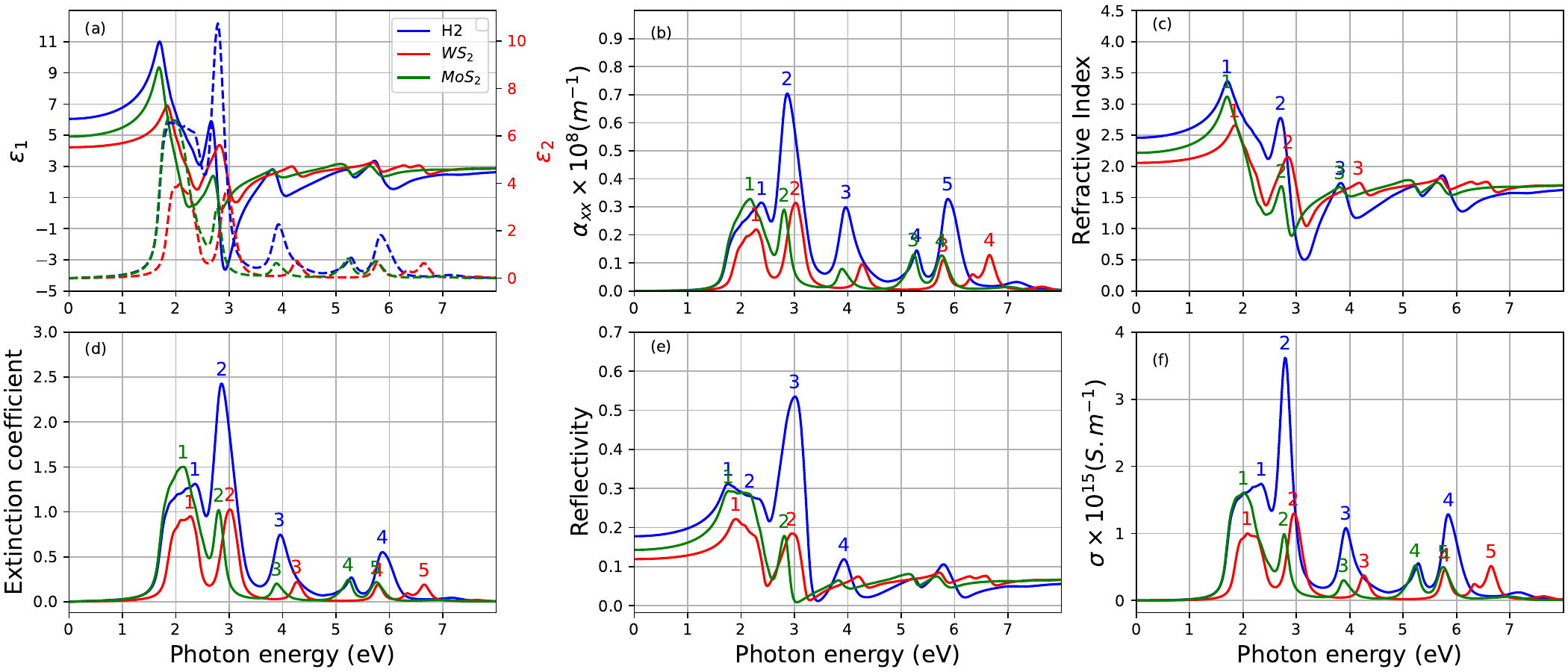} 
	\caption{(Colour online). Calculated optical properties for the H2 alloy (\ch{Mo_{0.875}W_{0.125}S2}, \SI{12.5}{\percent} W). (a) Real ($\epsilon_1$, solid line) and imaginary ($\epsilon_2$, dashed line) dielectric functions. Peaks in $\epsilon_2$ related to A and B transitions are blueshifted relative to H1, located near \SI{2.15}{\electronvolt} and \SI{2.38}{\electronvolt}. (b) Absorption coefficient ($\alpha$) showing a corresponding blueshift in the absorption edge and peaks. (c) Real refractive index ($n$) with dispersive features shifted to slightly higher energies. (d) Extinction coefficient ($k$) with A/B peaks near \SI{2.15}{\electronvolt} and \SI{2.38}{\electronvolt}. (e) Reflectivity ($R$) showing evolution consistent with changes in $n$ and $k$. (f) Optical conductivity ($\sigma$) with blueshifted primary peaks.}
    \label{fig:optic-h2}
\end{figure*}

\begin{figure*}[htpb]
	\centering
    \includegraphics[width=.9\linewidth]{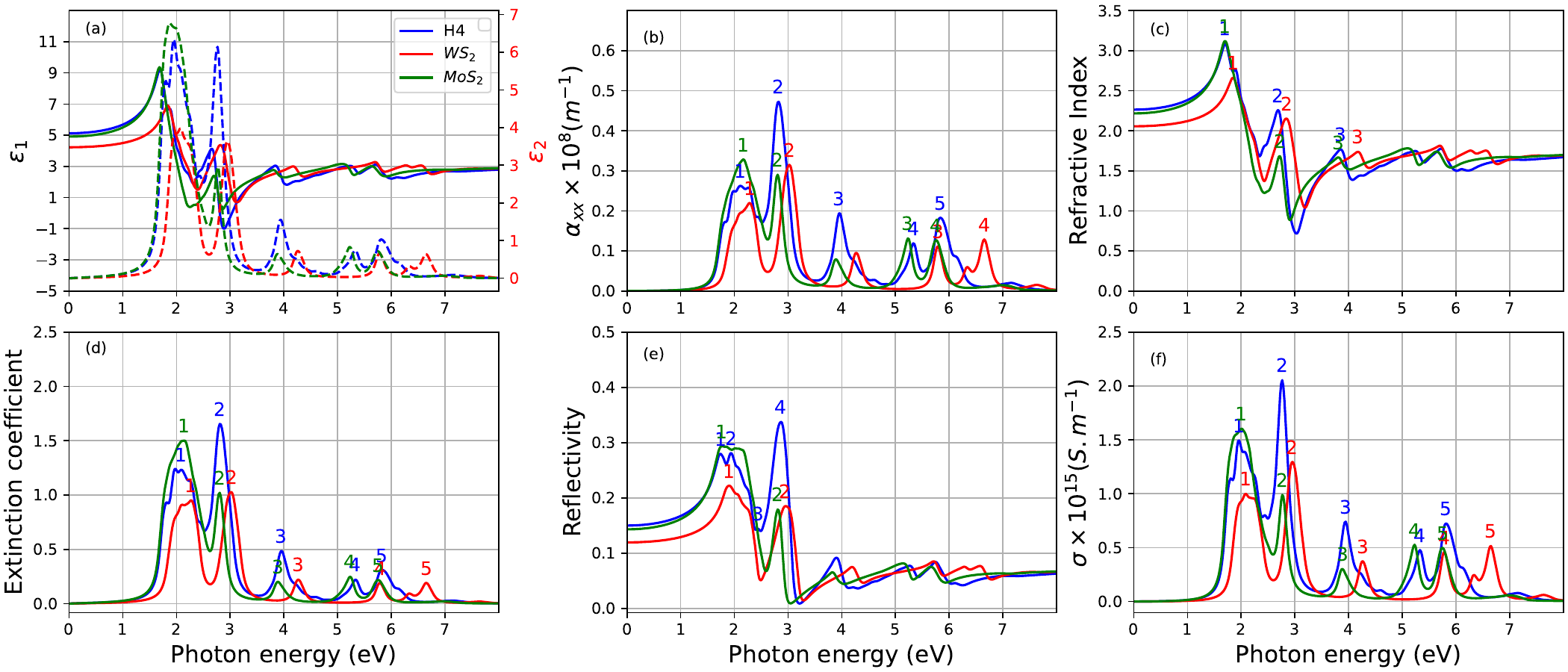}
	 \caption{(Colour online). Calculated optical properties for the H4 alloy (\ch{Mo_{0.75}W_{0.25}S2}, \SI{25.0}{\percent} W). (a) Real ($\epsilon_1$, solid line) and imaginary ($\epsilon_2$, dashed line) dielectric functions. A/B transition features in $\epsilon_2$ continue to blueshift, appearing near \SI{2.17}{\electronvolt} and \SI{2.40}{\electronvolt}. Note $E_{\text{peak}}$ from \Cref{tab:SummaryProperties} for this composition is \SI{2.08}{\electronvolt}. (b) Absorption coefficient ($\alpha$) reflecting the blueshifted absorption onset. (c) Real refractive index ($n$) shows evolving dispersive features. (d) Extinction coefficient ($k$) peaking near \SI{2.08}{\electronvolt}. (e) Reflectivity ($R$). (f) Optical conductivity ($\sigma$).}
    \label{fig:optic-h4}
\end{figure*}

\begin{figure*}[htpb]
	\centering
    \includegraphics[width=.9\linewidth]{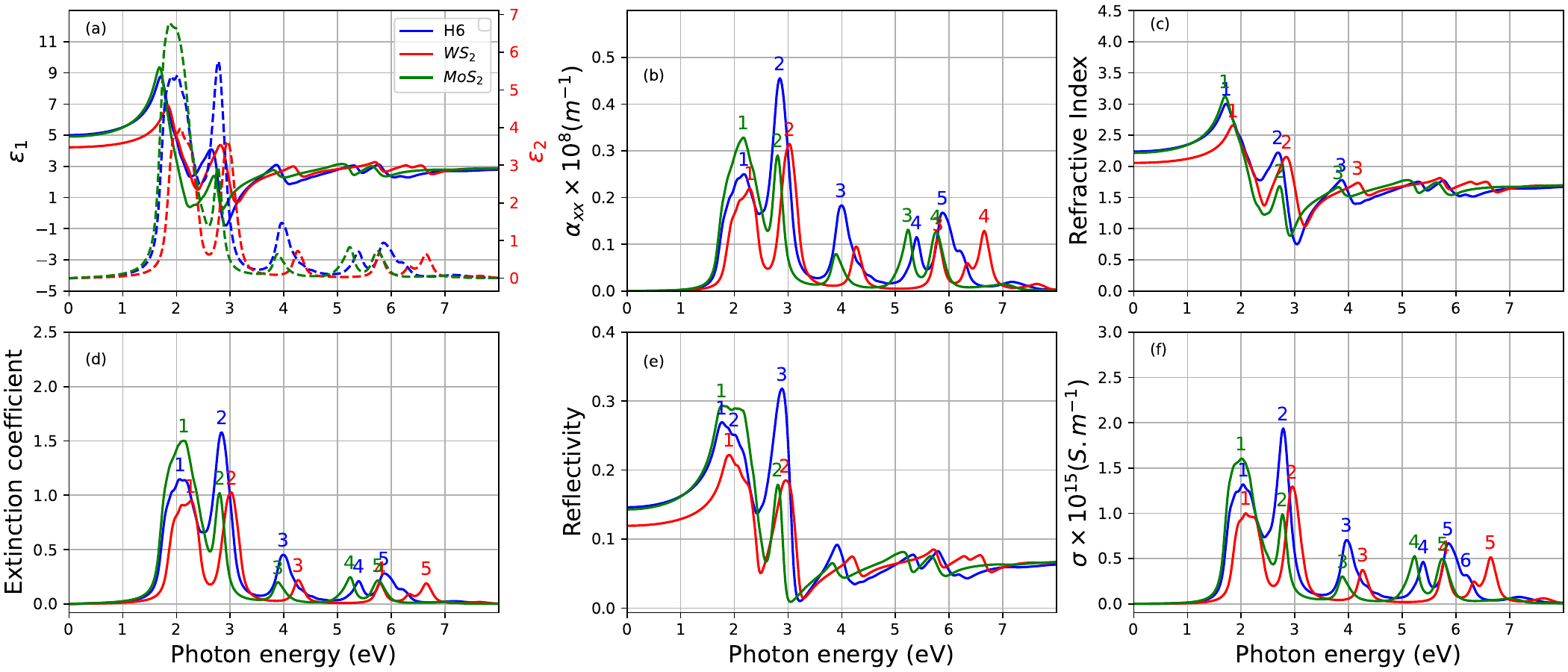}
	\caption{(Colour online). Calculated optical properties for the H6 alloy (\ch{Mo_{0.625}W_{0.375}S2}, \SI{37.5}{\percent} W). (a) Real ($\epsilon_1$, solid line) and imaginary ($\epsilon_2$, dashed line) dielectric functions. A/B features in $\epsilon_2$ are further blueshifted to approx. \SI{2.18}{\electronvolt} and \SI{2.42}{\electronvolt}. $E_{\text{peak}}$ from \Cref{tab:SummaryProperties} is \SI{2.15}{\electronvolt}. (b) Absorption coefficient ($\alpha$). (c) Real refractive index ($n$). (d) Extinction coefficient ($k$) peaking near \SI{2.15}{\electronvolt}. (e) Reflectivity ($R$). (f) Optical conductivity ($\sigma$).}
    \label{fig:optic-h6}
\end{figure*}

\begin{figure*}[htpb]
	\centering
    \includegraphics[width=.9\linewidth]{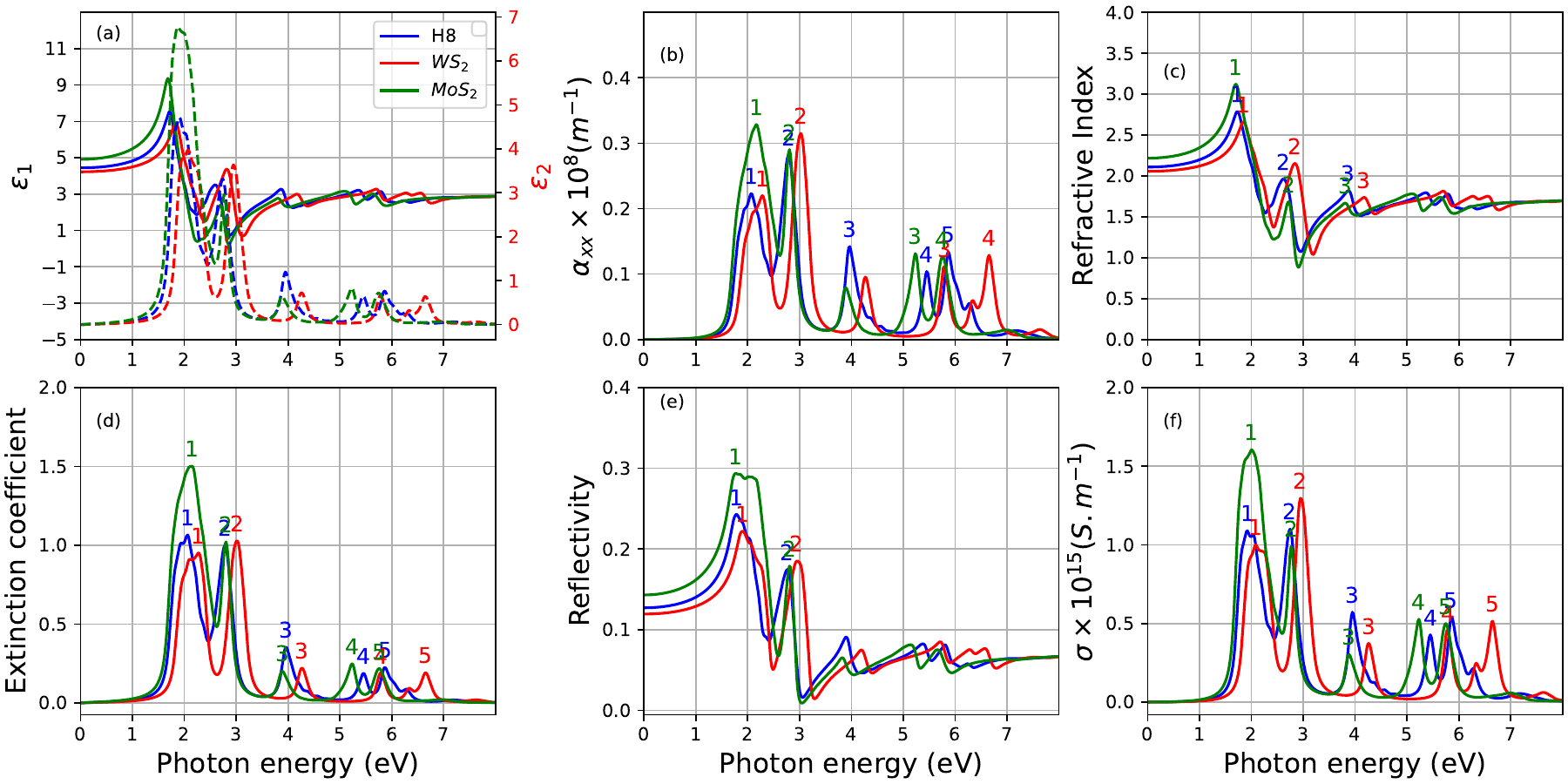}
	\caption{(Colour online). Calculated optical properties for the H8 alloy (\ch{Mo_{0.5}W_{0.5}S2}, \SI{50.0}{\percent} W). (a) Real ($\epsilon_1$, solid line) and imaginary ($\epsilon_2$, dashed line) dielectric functions. At midpoint composition, A/B features in $\epsilon_2$ are near \SI{2.20}{\electronvolt} and \SI{2.44}{\electronvolt}. $E_{\text{peak}}$ from \Cref{tab:SummaryProperties} is \SI{2.04}{\electronvolt}. (b) Absorption coefficient ($\alpha$). (c) Real refractive index ($n$). (d) Extinction coefficient ($k$) peaking near \SI{2.04}{\electronvolt}. (e) Reflectivity ($R$). (f) Optical conductivity ($\sigma$).}
    \label{fig:optic-h8}
\end{figure*}

\begin{figure*}[htpb]
	\centering
    \includegraphics[width=.9\linewidth]{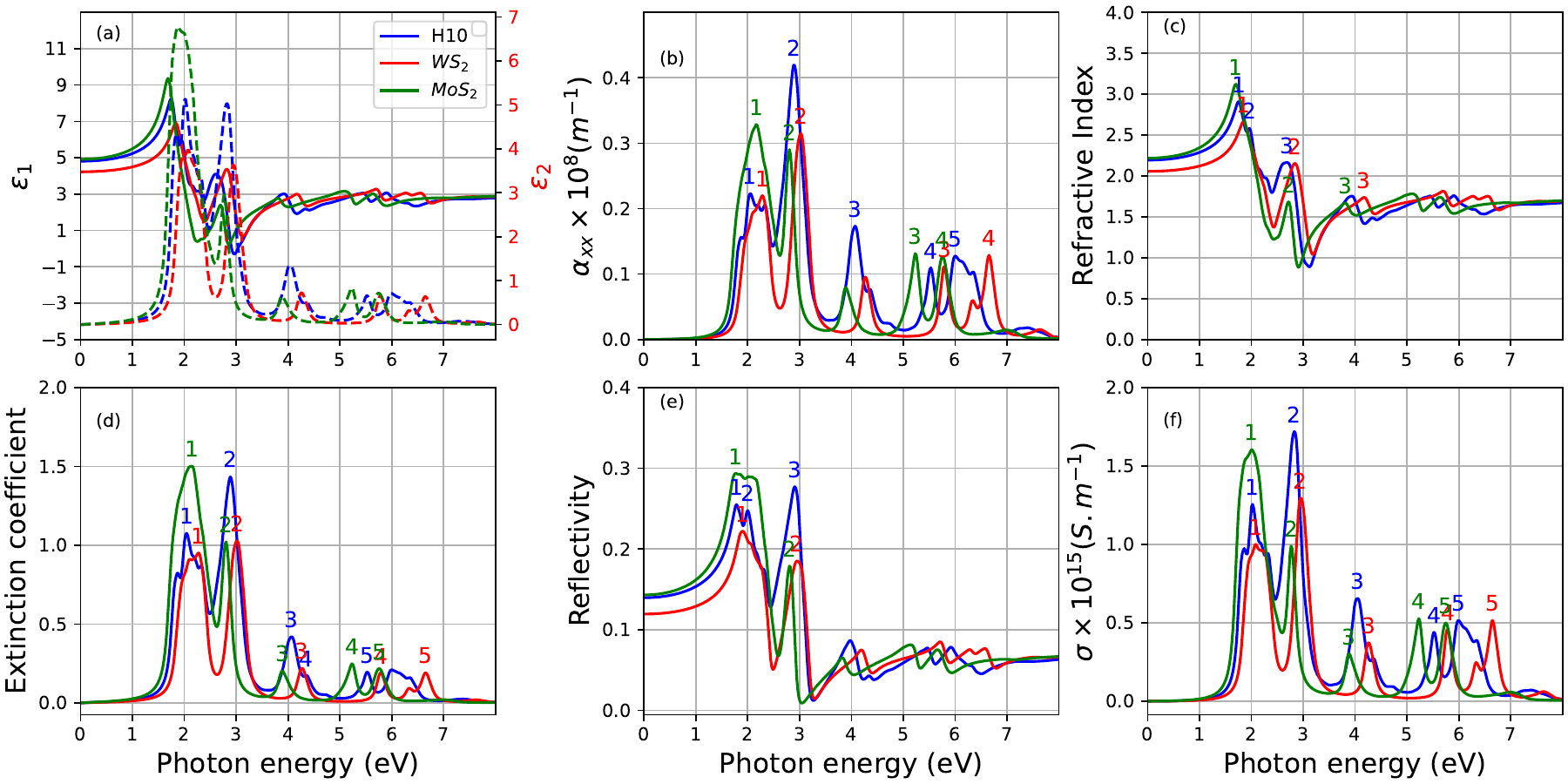}
	\caption{(Colour online). Calculated optical properties for the H10 alloy (\ch{Mo_{0.375}W_{0.625}S2}, \SI{62.5}{\percent} W). (a) Real ($\epsilon_1$, solid line) and imaginary ($\epsilon_2$, dashed line) dielectric functions. Continued blueshift places A/B features in $\epsilon_2$ around \SI{2.21}{\electronvolt} and \SI{2.46}{\electronvolt}. $E_{\text{peak}}$ from \Cref{tab:SummaryProperties} is \SI{2.04}{\electronvolt}. (b) Absorption coefficient ($\alpha$). (c) Real refractive index ($n$). (d) Extinction coefficient ($k$) peaking near \SI{2.04}{\electronvolt}. (e) Reflectivity ($R$). (f) Optical conductivity ($\sigma$).}
    \label{fig:optic-h10}
\end{figure*}

\begin{figure*}[htpb]
	\centering
     \includegraphics[width=.9\linewidth]{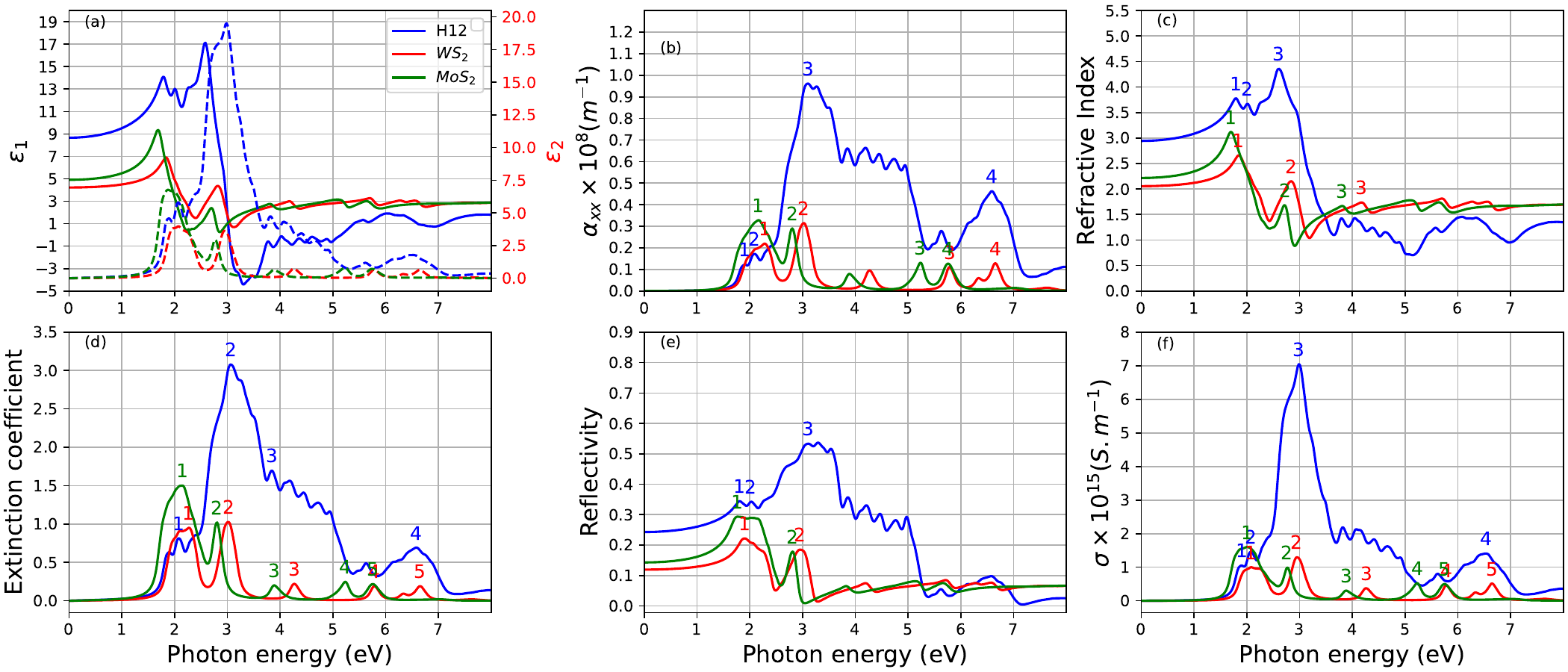} 
	\caption{(Colour online). Calculated optical properties for the H12 alloy (\ch{Mo_{0.25}W_{0.75}S2}, \SI{75.0}{\percent} W). (a) Real ($\epsilon_1$, solid line) and imaginary ($\epsilon_2$, dashed line) dielectric functions. A/B features in $\epsilon_2$ approach WS2 values, near \SI{2.23}{\electronvolt} and \SI{2.48}{\electronvolt}. $E_{\text{peak}}$ from \Cref{tab:SummaryProperties} is \SI{2.40}{\electronvolt}. (b) Absorption coefficient ($\alpha$). (c) Real refractive index ($n$). (d) Extinction coefficient ($k$) peaking near \SI{2.40}{\electronvolt}. (e) Reflectivity ($R$). (f) Optical conductivity ($\sigma$).}
    \label{fig:optic-h12}
\end{figure*}

\begin{figure*}[htpb]
	\centering
    \includegraphics[width=.9\linewidth]{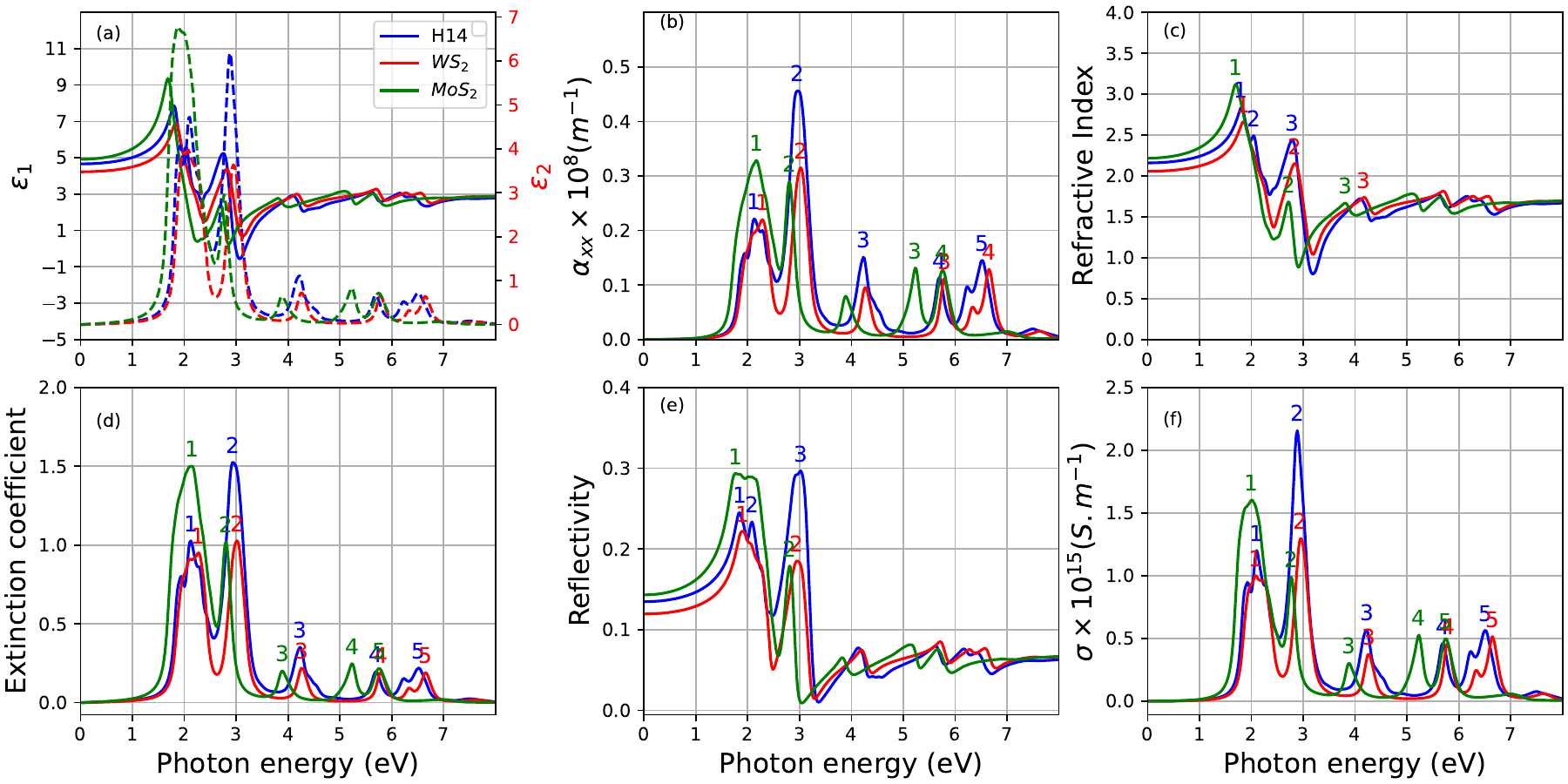}
	\caption{(Colour online). Calculated optical properties for the H14 alloy (\ch{Mo_{0.125}W_{0.875}S2}, \SI{87.5}{\percent} W). (a) Real ($\epsilon_1$, solid line) and imaginary ($\epsilon_2$, dashed line) dielectric functions. A/B features in $\epsilon_2$ are strongly blueshifted, located near \SI{2.25}{\electronvolt} and \SI{2.50}{\electronvolt}, approaching the pure \ch{WS2} limit. Note the $E_{\text{peak}}$ from \Cref{tab:SummaryProperties} for this composition is \SI{2.12}{\electronvolt}. (b) Absorption coefficient ($\alpha$). (c) Real refractive index ($n$). (d) Extinction coefficient ($k$) peaking near \SI{2.12}{\electronvolt}. (e) Reflectivity ($R$). (f) Optical conductivity ($\sigma$).}
    \label{fig:optic-h14}
\end{figure*}

Modifications to the electronic structure upon alloying, detailed in \Cref{subsec:Electronic}, directly influence the optical properties of the \ch{Mo_{1-x}W_xS2} monolayers. We investigated this optical response by calculating the complex frequency-dependent dielectric function $\varepsilon(\omega) = \varepsilon_1(\omega) + i\varepsilon_2(\omega)$ for each alloy composition using the methodology described in \Cref{sec:Methods}. Given the in-plane isotropy of the hexagonal structure, we focused on the response to light polarized parallel to the monolayer plane. From $\varepsilon(\omega)$, we subsequently derived key optical parameters, including the absorption coefficient $\alpha$, the refractive index $n$, the extinction coefficient $k$, the normal incidence reflectance $R$, and the optical conductivity $\sigma$. The evolution of these optical properties across the alloy series is presented in the spectral plots shown in \Crefrange{fig:h1-optic}{fig:optic-h14} (representing the H1 to H14 compositions as well as pristine \ch{MoS2} and \ch{WS2}) and summarized for key features in \Cref{tab:SummaryProperties}.

The calculated spectra reveal a rich optical response dominated by characteristics typical of monolayer TMDCs \cite{Wang2012}. The most prominent peaks observed near the absorption edge in the imaginary part of the dielectric function ($\epsilon_2$, panel (a) dashed lines in each figure) and consequently in the absorption coefficient ($\alpha$, panel (b)) correspond energetically to the interband transitions often associated with the A and B excitons observed experimentally. These arise fundamentally from direct transitions at the $K$-point involving the spin-orbit split highest valence bands and the lowest conduction band \cite{Komsa2013}. As anticipated from the RPA methodology which neglects electron-hole interactions (\Cref{sec:Methods}), the calculated energy for the primary absorption feature (A-peak) in pristine \ch{MoS2} (approx. \SI{2.14}{\electronvolt}, see \Cref{tab:SummaryProperties}) is significantly higher (by $\sim$\SIrange{0.2}{0.3}{\electronvolt}) than the experimentally observed A-exciton peak energy typically measured via photoluminescence or absorption ($\sim $\SIrange{1.8}{1.9}{\electronvolt}) \cite{Mak2010}. This difference highlights the strong excitonic effects present in these materials.

Despite the energy shift relative to the experiment, the trends observed in our calculations are highly relevant. As anticipated from the electronic structure analysis (\Cref{subsec:Electronic}), the positions of these primary spectral features (A and B transitions) exhibit a systematic blueshift with increasing concentration of \ch{W}, tracking the underlying increase in the direct bandgap. The onset of significant absorption, associated primarily with the A feature, progressively shifts from approximately \SI{1.8}{\electronvolt} in \ch{MoS2}-rich alloys towards \SI{2.0}{\electronvolt} as the alloys become \ch{WS2}-rich. Quantitatively, the calculated peak associated with the A transition ($E_{\text{peak}}$ in \Cref{tab:SummaryProperties}) shifts from \SI{2.14}{eV} ($x=0$) to \SI{2.26}{\electronvolt} ($x=1$). This energy shift corresponds to a wavelength tuning range of approximately \SI{30}{\nm} for this calculated interband peak. Although the absolute experimental peak positions will be reduced due to exciton binding, the relative compositional tuning range observed in PL experiments ($\sim $\SI{0.18}{\electronvolt} \cite{Liu2017}) is well captured by our bandgap calculations, suggesting a similar substantial tunability for the actual optical response.

While the dominant trend for the main absorption features is this systematic blueshift, mirroring the bandgap trend, the evolution of higher-energy optical features (particularly in the range \SIrange{2.5}{8}{\electronvolt}) is more complex. These features originate from transitions involving electronic states deeper in the valence band or higher in the conduction band, often localized near different points in the Brillouin zone (e.g., near $\Gamma$ or M) \cite{Komsa2013, Tang2018}. Their energies and intensities are sensitive to the joint density of states (JDOS) and specific band nesting conditions, which can evolve non-monotonically or non-linearly as the electronic structure gradually changes upon alloying. Despite this complexity, it is noteworthy that the overall spectral shape evolves smoothly across the composition range without abrupt changes (such as the sudden appearance/disappearance of major peaks), suggesting that no dramatic electronic phase transitions occur. The spectra gradually morph from the characteristic response of \ch{MoS2} to that of \ch{WS2}. \textbf{Notably, for very low tungsten concentrations (H1, $x=0.0625$ and H2, $x=0.125$), the calculated optical response, particularly the absorption and extinction coefficients in the \SIrange{2.27}{6.3}{\electronvolt} range, and the refractive index between \SIrange{0}{3.1}{\electronvolt}, surpasses that of both pristine \ch{MoS2} and \ch{WS2} (see \Cref{fig:h1-optic,fig:optic-h2}). This suggests a potentially non-trivial enhancement of optical activity due to specific Mo-W interactions at low alloying levels, which could be interesting for applications requiring higher optical interaction in these energy windows. Similarly, the reflectivity and optical conductivity for these two compositions also appear to be enhanced in specific energy ranges compared to the parent compounds.} As the percentage of W increases further beyond these initial enhancements, the optical response of the alloys progressively approaches that of pure \ch{WS2}. \textbf{Interestingly, another distinct behavior is observed at $x=0.75$ (H12, \Cref{fig:optic-h12}), where the optical response, particularly for absorption and extinction coefficients in the \SIrange{2.27}{8}{\electronvolt} range, again appears to surpass that of both parent compounds. This non-monotonic behavior at specific compositions merits further investigation, as it might indicate complex electronic interactions or structural relaxations that favorably modify the joint density of states for optical transitions.}

This systematic tuning of the absorption edge via alloying presents a significant advantage. The controllable shift in the primary absorption onset, reflecting the bandgap tunability, spans a wavelength range of approximately \SI{100}{\nm} across the full composition range ($x=0$ to $x=1$). This tunability significantly exceeds that typically reported for simple vertically stacked \ch{MoS2}/\ch{WS2} heterostructures (often limited to $\Delta \lambda \approx \SI{40}{\nm}$ due to fixed band alignments), suggesting that these alloys may offer greater flexibility for applications requiring adjustable wavelength selectivity or broadband photodetection integrated into a single material layer.

\begin{figure*}[htpb]
	\centering
    \includegraphics[width=.9\linewidth]{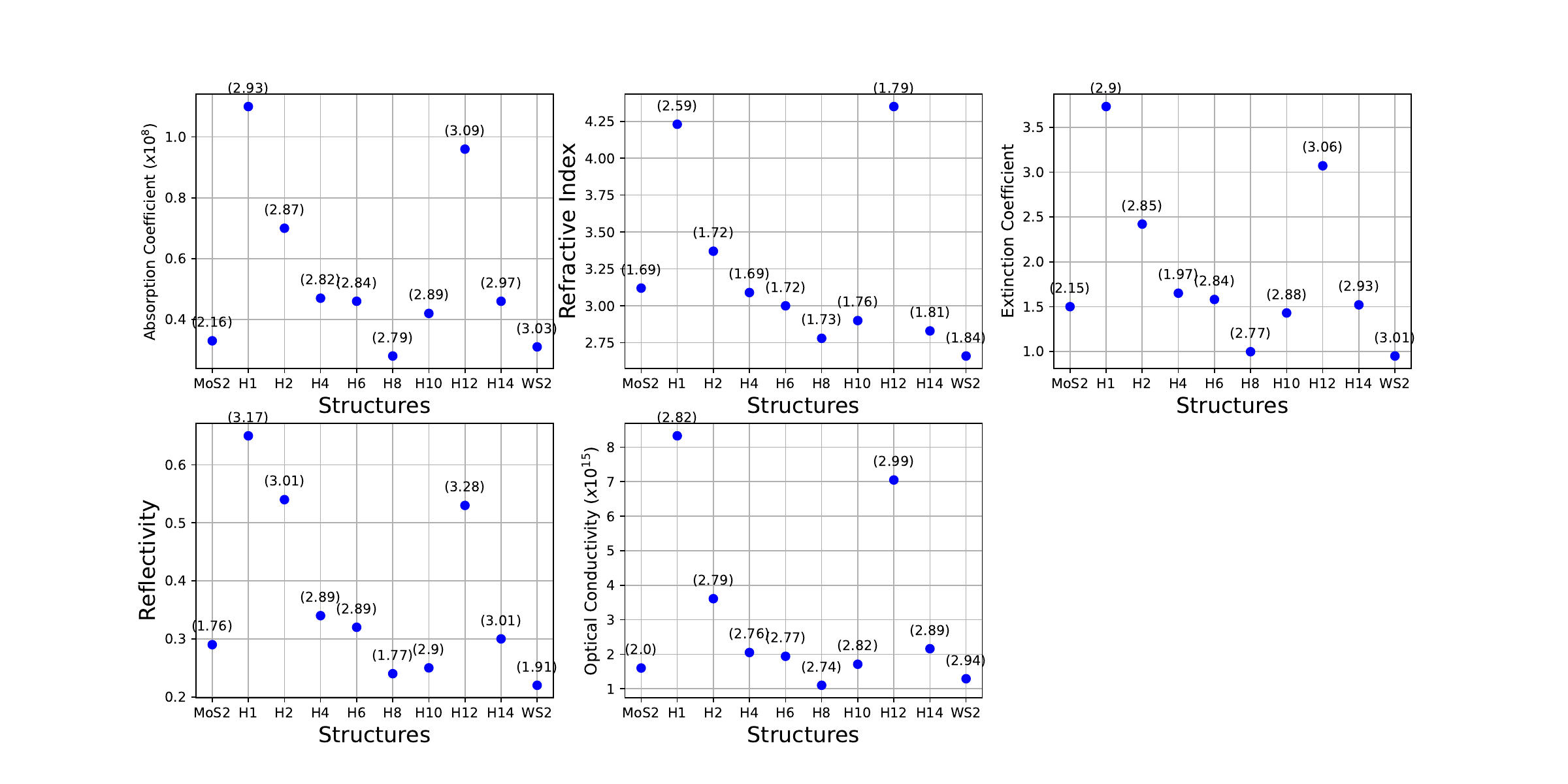}
	\caption{(Colour online). Evolution of the maxima of different optical quantities as a function of composition. Values in parentheses indicate the photon energy (\unit{\electronvolt}) at which the respective maximum occurs.}
    \label{fig:Evolution-optical}
\end{figure*}

Finally, the derived optical constants related to refraction and reflection also exhibit predictable compositional dependence. The real part of the refractive index ($n$, panel (c) of the figures) displays the characteristic dispersive features (peaks and troughs) associated with the main absorption bands, consistent with the Kramers-Kronig relationship. The extinction coefficient ($k$, panel (d)) closely mirrors the behavior of the absorption coefficient $\alpha$, as expected. The normal incidence reflectance ($R$, panel (e)) and optical conductivity ($\sigma$, panel (f)) also show features corresponding to the main transitions, which evolve systematically with composition. The evolution of maxima for different optical quantities with varying compositions is illustrated in \Cref{fig:Evolution-optical}. The ability to engineer not only the absorption but also the refractive index and reflectance across the visible and near-UV spectrum further enhances the potential of \ch{Mo_{1-x}W_xS2} alloys for designing specific optical components, optimizing light management in photovoltaic devices, or developing color-tuned light-emitting diodes. However, achieving sufficient absorption or emission efficiency in practical devices remains a challenge due to the inherent atomic thinness of the monolayer, which requires further device engineering strategies. Our comprehensive analysis up to \SI{8}{\electronvolt} provides a valuable baseline understanding of how alloying controls the complete optical response in this 2D system.

\begin{table*}[htpb]
    \centering
    \caption{Summary of key calculated properties for \ch{Mo_{1-x}W_xS2} monolayer alloys. $E_g$ is the direct Kohn-Sham bandgap at K. $a$ is the average lattice parameter in the plane (derived from M-M distances). S-S is the average layer thickness. $E_b$ is the formation energy per atom (Eq. \eqref{eq:FormationEnergy}). $n \text{@} \SI{1.5}{\electronvolt}$ is the refractive index at $\SI{1.5}{\electronvolt}$. $k_{\text{peak}}$ is the peak extinction coefficient near the A exciton region, occurring at energy $E_{\text{peak}}$. Missing data for $k_{\text{peak}}$ are denoted by '--' and typically indicate that the peak was not clearly resolved or data extraction was problematic for those specific compositions.}
    \label{tab:SummaryProperties}
    \sisetup{round-mode=places, round-precision=3}
    \begin{tabular}{@{} l
                      S[table-format=1.4] 
                      S[table-format=1.3] 
                      S[table-format=1.3] 
                      S[table-format=1.3] 
                      S[table-format=-1.3] 
                      S[table-format=1.2] 
                      c @{}} 
        \toprule
        Structure & {$x$} & {$E_g$} & {$a$} & {S-S} & {$E_b$} & {$n$} & {$k_{\text{peak}}$} \\
                  & {(W frac.)} & {(\unit{\electronvolt})} & {(\unit{\angstrom})} & {(\unit{\angstrom})} & {(\unit{eV/atom})} & {@ $\SI{1.5}{\electronvolt}$} & {($E_{\text{peak}}$ in \unit{\electronvolt})} \\
        \midrule
        \ch{MoS2}              & 0.0    & 1.696 & 3.180 & 3.129 & -6.910 & 2.72 & 1.50 (2.14) \\
        H1 (\SI{6.25}{\percent} W) & 0.0625 & 1.700 & 3.184 & 3.127 & -6.970 & 3.59 & 1.31 (2.36) \\
        H2 (\SI{12.5}{\percent} W) & 0.125  & 1.709 & 3.183 & 3.128 & -6.980 & 2.99 & 1.31 (2.36) \\
        H4 (\SI{25.0}{\percent} W) & 0.25   & 1.723 & 3.180 & 3.129 & -7.040 & 2.72 & 1.23 (2.08) \\
        H6 (\SI{37.5}{\percent} W) & 0.375  & 1.741 & 3.183 & 3.132 & -7.070 & 2.65 & 1.12 (2.15) \\
        H8 (\SI{50.0}{\percent} W) & 0.5    & 1.737 & 3.180 & 3.133 & -7.130 & 2.44 & 1.05 (2.04) \\
        H10 (\SI{62.5}{\percent} W)& 0.625  & 1.760 & 3.180 & 3.132 & -7.180 & 2.53 & 1.07 (2.04) \\
        H12 (\SI{75.0}{\percent} W)& 0.75   & 1.794 & 3.182 & 3.134 & -7.240 & 3.37 & 0.86 (2.40) \\
        H14 (\SI{87.5}{\percent} W)& 0.875  & 1.818 & 3.180 & 3.136 & -7.280 & 2.43 & 1.03 (2.12) \\
        \ch{WS2}               & 1.0    & 1.858 & 3.180 & 3.147 & -7.370 & 2.28 & 1.03 (2.26) \\
        \bottomrule
    \end{tabular}
\end{table*}

\section{\label{sec:Conclusion}Conclusion} 

This work presented the first comprehensive investigation of density functional theory (to our knowledge) that systematically maps the structural, electronic, and optical properties of monolayer \ch{Mo_{1-x}W_xS2} alloys across the entire compositional spectrum from $x=0$ (pure \ch{MoS2}) to $x=1$ (pure \ch{WS2}). By addressing this critical gap in the literature, which often focused on limited compositions, our results confirm the thermodynamic stability of these alloys and show that they consistently retain the hexagonal 2H structure with minimal perturbation upon alloying. 

A key finding of this study is the preservation of a direct electronic bandgap located at the $K$-point of the Brillouin zone in all alloy compositions ($x=0 \text{ to } 1$). Furthermore, the magnitude of this direct gap demonstrates continuous tunability, increasing near-monotonically from approximately \SI{1.696}{\electronvolt} for \ch{MoS2} to \SI{1.858}{\electronvolt} for \ch{WS2}. This controllable bandgap engineering is a key attribute that is essential for tailoring the electronic behavior of the material to meet specific optoelectronic device requirements, enabling, for example, the design of photodetectors with spectral sensitivity spanning the red to near-infrared range (\SIrange{\sim 650}{\sim 720}{\nm}) or light emitters with customizable colors within this spectrum.

Congruent with these electronic structure modifications, the calculated optical properties, derived from the complex dielectric function, exhibit a systematic evolution. The dominant optical features, corresponding to interband transitions associated with the A and B excitons, display a characteristic blueshift that mirrors the underlying bandgap increase. The predictable compositional dependence of the absorption coefficient, refractive index, and extinction coefficient within the evaluated visible and near-UV range (\SIrange{1.5}{8}{\electronvolt}) provides the foundation for precisely tailoring light-matter interactions in wavelength-specific applications.

Taken together, our findings establish monolayer \ch{Mo_{1-x}W_xS2} alloys as an electronically and optically versatile platform for bandgap engineering within the family of 2D transition metal dichalcogenides. By providing a continuous picture of property evolution, this work highlights several key attributes: confirmed thermodynamic stability, structural integrity, the critical retention of a continuously tunable direct electronic gap (\SIrange{1.696}{1.858}{\electronvolt}), and a systematically adjustable optical response across the visible spectrum. Notably, the demonstrated tunability of primary absorption features corresponds to a wavelength range of approximately \SI{100}{\nm} ($\sim$ \SIrange{550}{650}{\nm} based on $E_{\text{peak}}$ shifts), offering a significantly broader continuous tuning capability compared to the typical $\sim$ \SI{40}{\nm} range reported for vertically stacked \ch{MoS2}/\ch{WS2} heterostructures, thus highlighting a distinct advantage of the alloying strategy for certain applications. These characteristics strongly support the suitability of \ch{Mo_{1-x}W_xS2} alloys for realizing next-generation optoelectronic devices, including wavelength-tunable photodetectors, customized light emitters and components for flexible optoelectronics \cite{Aftab2023}, positioning these alloys as strong candidates for integration into practical technologies based on their tunable electronic/optical properties and inherent 2D nature.

While this study provides essential insights into the fundamental properties of \ch{Mo_{1-x}W_xS2} alloys, we acknowledge the limitations of standard DFT in capturing strong excitonic effects, which significantly influence experimental optical spectra. Future investigations employing higher-level theories, such as calculations using the GW approximation plus the Bethe-Salpeter equation (GW-BSE) \cite{Komsa2013}, are warranted to accurately predict exciton binding energies and refine absolute optical transition energies across the alloy series. Complementary experimental validation will also remain essential to fully harness the potential of these materials in practical devices. Overall, this work offers a detailed computational foundation and roadmap, contributing to the rational design and optimization of \ch{MoS2}/\ch{WS2} alloy-based nanomaterials for emerging electronic and optical technologies.

\section*{\label{sec:acknowledgements}Acknowledgements} 

The authors acknowledge the Centre for High Performance Computing (CHPC), South Africa, for providing computational resources for this research project.

\section*{\label{sec:Data availability}Data availability statement}
The data that support the findings of this study are available upon reasonable request from the authors.


%

\end{document}